

\input phyzzx

\catcode`@=11
\newtoks\KUNS
\newtoks\HETH
\newtoks\monthyear
\Pubnum={KUNS~\the\KUNS\cr HE(TH)~\the\HETH}
\monthyear={December, 1992}
\def\p@bblock{\begingroup \tabskip=\hsize minus \hsize
   \baselineskip=1.5\ht\strutbox \topspace-2\baselineskip
   \halign to\hsize{\strut ##\hfil\tabskip=0pt\crcr
   \the\Pubnum\cr hep-ph/9212317\cr \the\monthyear\cr }\endgroup}
\def\bftitlestyle#1{\par\begingroup \titleparagraphs
     \iftwelv@\fourteenpoint\else\twelvepoint\fi
   \noindent {\bf #1}\par\endgroup }
\def\title#1{\vskip\frontpageskip \bftitlestyle{#1} \vskip\headskip }
%
%

%
%
\paperfootline={\hss\iffrontpage\else\ifp@genum%
                \tenrm --\thinspace\folio\thinspace --\hss\fi\fi}
\footline=\paperfootline
%
%

%
\def\journal#1&#2(#3){\begingroup \let\journal=\dummyj@urnal
    \unskip, \sl #1\unskip~\bf\ignorespaces #2\rm
    (\afterassignment\j@ur \count255=#3) \endgroup\ignorespaces }
\def\andjournal#1&#2(#3){\begingroup \let\journal=\dummyj@urnal
    \sl #1\unskip~\bf\ignorespaces #2\rm
    (\afterassignment\j@ur \count255=#3) \endgroup\ignorespaces }
\def\andvol&#1(#2){\begingroup \let\journal=\dummyj@urnal
    \bf\ignorespaces #1\rm
    (\afterassignment\j@ur \count255=#2) \endgroup\ignorespaces }
%

%




\KUNS={1174}     
\HETH={92/14}   

\titlepage

\title{A Signal for Technicolor ?}

\author{Masako Bando}

\address{Physics Division, Aichi University, Miyoshi,
                                             Aichi 470-02, Japan}
\andauthor{Nobuhiro Maekawa }

\address{Department of Physics, Kyoto University, Kyoto 606, Japan}



\abstract{
We propose an interpretation for the
($l\bar l\gamma \gamma ,\ M_{\gamma \gamma }=60{\rm GeV}$)
events, which have recently been  reported by L3 group at LEP.
This may be a first signal of `Technicolor' theory.
}

\endpage          

\REF\La{
 L3 Collaboration, CERN Report CERN-PPE/92-152(1992).
}

\REF\BDHR{
 V.~Barger, N.G.~Deshpande, J.L.~Hewett and T.G.~Rizzo,
                Preprint ANL-HEP-PR-92-102.
}

\REF\KG{
 D.B~.~Kaplan and H.~Georgi
               \journal Phys. Lett. &136B (84) 183.
}

\REF\GKD{
 D.B~.~Kaplan, H.~Georgi and S.~Dimopoulos
               \journal Phys. Lett. &136B (84) 187.
}

\REF\WZ{
 J.~Wess and B.~Zumino
                \journal Phys. Lett. &37B (71) 95.
}

\REF\FS{
 E.~Farhi and L.~Susskind
  \journal Phys. Rep. &74 (92) 277.
}

\overfullrule=0pt

The L3 group of LEP experiments has reported $l^+l^-\gamma \gamma $
events with
$M_{\gamma \gamma }\simeq 60{\rm GeV}$.
\refmark\La
 There has been pointed out that this would be
a signal for $Z\rightarrow (Z^*\rightarrow l^+l^-)
+(h^0\rightarrow \gamma \gamma )$, where $Z^*$ is the
off-shell $Z$ boson and $h^0$ is
the on-shell lightest
CP-even neutral Higgs which completely decouples from the ordinary
quarks and leptons but has couplings to the $W$, $Z$ gauge bosons and
Higgs bosons( a separate Higgs scenario).\refmark\BDHR
Let us denote the virtual (off-shell) particles by attaching $*$
to the particle names generically in this paper.

However this interpretation has the following serious difficulties:

\item{1)} Since $B(Z\rightarrow \nu \bar \nu )/B(Z\rightarrow e
\bar e)\simeq 6$,
the L3 group should
have observed approximately 3 times as many $2\gamma $ plus missing
energy
events as the sum of $e\bar e\gamma \gamma $ and
$\mu \bar \mu \gamma \gamma $ events
(we neglect the event $\tau \bar \tau $ because of harder
experimental cut).
However no such event has been found within the range
$M_{\gamma \gamma }>10{\rm GeV}$
at LEP.

\item{2)} The separate Higgs
 $h^0$
does not have a direct photon coupling and so
 $(h^0\rightarrow \gamma \gamma )$ can proceed only via one
loop diagrams of
charged
particles. The only candidates for these here are charged Higgs,
which demands that
 $h^0$ strongly couples to
 $H^\pm $. Then in order to suppress
 $h^0$ decays to the ordinary lepton/quark pairs,
the parameters of the scalar Higgs interactions should be fine tuned
to have maximum charged Higgs loop contributions.

To solve these difficulties, we should notice the following facts.

Let
 $X$ be the new particle with its mass
 $M_{\gamma \gamma }\simeq 60{\rm GeV}$ which decays mainly into
 $\gamma \gamma $  and
 $Y$ be the associated virtual particle whose main decay modes
are charged lepton pairs and not neutrino pair (Fig.1).

First note that
        $Y$ has to couple to charged lepton pairs,
 $e \bar e$ or
 $\mu  \bar \mu $, but not
 $\nu  \bar \nu $.
So $Y$ is not
 $Z$.
Then what is $X$ which forbids the tree vertex
 $XZZ$?
The answer is that
 $X$ is a  pseudoscalar (i.e., CP-odd) particle $P$.
CP invariance forbids the tree vertex $PZZ$
(CP allows $P\epsilon ^{\mu \nu \rho \sigma }\partial _\mu Z_\nu
\partial _\rho Z_\sigma $ term,
but this is a non-renormalizable (dimension 5) operator).
This is one important point of this paper.
To see this we would like to remind the usual scalar
interaction;
$$
  |(\partial _\mu -ig_ZZ_\mu )\phi _0|^2={1\over 2}(\partial _\mu S)^2
                                 +{1\over 2}(\partial _\mu P)^2
        -{1\over 2}g_Z^2Z_\mu Z^\mu (S^2+P^2)-ig_ZZ^\mu
                (\partial _\mu S\cdot P-S\partial_\mu  P),
$$
where $\phi _0=(S+iP)/\sqrt{2}$ is a neutral scalar and $g_Z$ is a
gauge coupling to $\phi _0$.
And $S$ is CP even scalar particle, and $P$ is a CP odd one.
{}From the above equation,
we can find that $P$ is a good candidate for $X$
because there is no vertex $PZZ$ ( while the $SZZ$ vertex would
exist generally owing to $\VEV{S}\not=0$).
Once we regard $P$ as $X$,
 we can take $S$ as $Y$, because there is a vertex $ZPS$.

In this way it turns out that a favorable possibility is that
 $X$ is a pseudoscalar particle and
 $Y$ is a scalar particle.

We propose here the following simple Technicolor(TC) scenario
which can evade  the difficulties of the separate Higgs interpretation;
$Z\rightarrow (S_T^*\rightarrow l\bar l)+(P_T\rightarrow \gamma
\gamma )$ (Fig.2),
where $S_T^*$ is a virtual technibound state (neutral scalar
TC bound state) and
 $P_T$ is a on-shell neutral techni pion
(pseudo Nambu-Goldstone boson (PGB)).
Namely $X=P_T$, $Y=S_T$.
And we assume
that the techni fermions (TF) play
only a role for breaking electroweak symmetry,
 but that they never give the masses of leptons and quarks.
It is well known that generally the Higgs fields
play dual roles in the standard model, generating the electroweak
symmetry breakdown to yield massive $W$ and $Z$
bosons on the one hand, and giving the masses of
quarks and leptons on the other hand.
TC does a beautiful explanation for electroweak symmetry
breaking but for the latter it still has various problems to be solved.
We here would like to separate the above roles in TC theory, and
introduce only TC interaction
{\it without extended TC(ETC)} interactions( a Separate TC).
This is another important point.
This kind of scenario (although in a bit different way)
  was first proposed by
Georgi and Kaplan named as `oblique hypercolor'' theory.
\refmark{\KG, \GKD}
This  attitude seems to be favorable for those who want to
build more a consistent and realistic models
(due to so called FCNC problem ).
 Under the condition of no ETC interaction,
\foot{It may be possible to introduce ETC like interactions
so far as we keep their strength within the limit of experimental
constraint of the decay width $\Gamma (X\rightarrow all)
\sim \Gamma (X\rightarrow \gamma \gamma )$. However
we here assume complete decoupled TC model for simplicity.}
 $P_T\rightarrow 2\gamma $ decay is generated
via TF loops just like $\pi ^0\rightarrow 2\gamma $ decays.

 $S_T$ can couple to gauge bosons via the TF loops
which yields
 $S_T^*\rightarrow \gamma ^*\rightarrow l\bar l$,
 $S_T^*\rightarrow \gamma \gamma $ and
 $S_T^*\rightarrow Z^*\rightarrow l\bar l,\nu \bar \nu $,
and so on.
The L3 data implies that the first process, $S_T^*\rightarrow
\gamma ^*\rightarrow l\bar l$,
 must dominate. Here we analyze under this  assumption.
 Then $S_T$ can decay mainly into charged lepton pairs with the ratio
$$
       {B(S_T\rightarrow \nu \bar \nu )\over B(S_T\rightarrow e
\bar e)}\ll 1,
$$
which is nicely in favor to the fact that $\gamma \gamma +$missing
energy process
are not observed.
Similarly, we obtain,
$$
R(S_T)\equiv {B(S_T\rightarrow q\bar q)\over B(S_T\rightarrow e\bar e)}
       ={B(\gamma \rightarrow q\bar q)\over B(\gamma \rightarrow e
\bar e)}
       ={11\over 3}.
$$
This ratio is to be compared with the value of
the Separate Higgs model (Z boson is mediated):
$R(h^0)\equiv {B(Z\rightarrow q\bar q)/B(Z\rightarrow e\bar e)}=20.85$
( This is experimental value measured at LEP ).
Thus our model predicts less cross section for $\gamma \gamma +2\
jets$ mode.
More accurate experimental data is desired to clarify this point.

The question is whether $S_T^*\rightarrow \gamma $ process dominates
or not.
To answer this we have to solve the model dynamically.
The QCD analogous discussion teaches us that $S$ decays mainly into
 $\gamma \gamma $, unfortunately. On the other hand, $\rho $
is known to
decay into lepton pairs and
C invariance forbids $\gamma \gamma $ decay.
So
if
 $Y$ is a vector state $V_T$ (like $\rho $ meson),
it couples directly to one photon.
This replacement may  solve all the above difficulties.

Finally we comment on the mass of PGBs.
If one takes the usual TC one family model,
 for example, pseudo
goldstone bosons(PGB's) are produced, which acquire masses via
the chiral-symmetry violating gauge interactions (other than TC).
The mass of the lightest PGB $P_T$ is estimated via Dashen formula,
\refmark\FS
$$
0<m_{P_T}<100{\rm GeV}.
$$
which is to be compared with the invariant mass
$M_{\gamma \gamma }\simeq 60{\rm GeV}$.

In conclusion we here recapulate the main points of our results.
\item{a)} The new particle with $M_{\gamma \gamma }=60{\rm GeV}$
observed by L3
group
should be a CP-odd particle,
because the tree vertex $ZZP$ is forbidden by CP invariance.
The point is that the existence of $Z^*\rightarrow \nu \bar \nu
\gamma \gamma $ vertex
contradicts the fact
that $\nu  \bar \nu \gamma \gamma $ event has not been observed yet
at LEP.

\item{b)} The L3 event prefers to be interpreted by
the Separate TC(STC) theory in which techni-mesons do not have
direct couplings to ordinary quarks and leptons.
This gives a natural interpretation why the main decay
mode of the new particle is $\gamma \gamma $.

There are a variety of possibilities as the candidates for such
 a CP-odd particle and for such separate TC fermions,
which will be investigated in details in the forthcoming papers.

We would like to thank T.Kugo and M.G.Mitchard for their reading
our manuscript and for their valuable
comments and discussions.
\refout

\vskip 1cm

{\bf FIGURE CAPTION}

\item{{\bf Fig.1}} \ A Fenynman diagarm of the L3 events
($Z\rightarrow (X\rightarrow \gamma \gamma )+(Y\rightarrow l\bar l)$).

\item{{\bf Fig.2}} \ A Feynman diagram contributing
to $l\bar l\gamma \gamma $ process,
which is intermediated by techni vector $V_T$ or by techni scalar
$S_T$ via TF loop.

\bye